\documentclass[a4paper]{jpconf}
\usepackage{graphicx}
\usepackage{amssymb}
\usepackage{cite}
\usepackage{color}
\begin{document}
\title{Generation of soliton bubbles in a sine-Gordon system with localised inhomogeneities}

\author{Juan F. Mar\'in}

\address{Instituto de F\'isica, Pontificia Universidad Cat\'olica de Valpara\'iso, Avenida Brasil,
Valpara\'iso, Casilla 2950, Chile.}

\ead{juanfmarinm@gmail.com}

\begin{abstract}
Nonlinear wave propagation plays a crucial role in the functioning of many physical and biophysical systems. In the propagation regime,  disturbances
due to the presence of local external perturbations, such as localised defects or boundary interphase walls have gained great attention. In this
article, the complex phenomena that occur when sine-Gordon line solitons collide with localised inhomogeneities are investigated. By a one-dimensional
theory, it is shown that internal modes of two-dimensional sine-Gordon solitons can be activated depending on the topological properties of the
inhomogeneities. Shape mode instabilities cause the formation of bubble-like and drop-like structures for both stationary and travelling line
solitons. It is shown that such structures are formed and stabilised by arrays of localised inhomogeneities distributed in space.
Implications of
the observed phenomena in physical and biological systems are discussed.
\end{abstract}

\section{Introduction}
\label{Sec:Introduction}

The propagation of nonlinear waves in inhomogeneous media has captured much recent interest in physics and biophysics \cite{Ivancevic2013}. Nonlinear
pulse excitations regulate the functioning in many physical models of biological systems, such as the Purkinje's fibres of the cardiac muscles
\cite{Aslanidi1999, Mornev2000}, neural fibres \cite{Scott1975}, DNA chains \cite{Polozov1988, Yakushevich2004} and muscular networks
\cite{Gojkovic2016}. The study of the effects that can be produced on these regulatory signals by the presence of spatial inhomogeneities, such as
localised injuries and necrotic sites, has gained major interest in recent works \cite{Mornev2000, Grinevich2015, Grinevich2016}. For instance,
injuries in the heart muscular tissue may produce annihilation, transmission or reflection of incident waves, depending on the geometrical size and
shape of the defects. These disturbances may produce a disorder in the rhythmic contractions of the muscular walls,
leading to cardiac arrhythmias and different pathologies \cite{Mornev2000}. A similar situation occurs in the propagation of topological
one-dimensional solitons in DNA chains, which has been considered as a suitable model to explain the formation of open states or {\it bubbles} in the
double helix \cite{Grinevich2015}. This bubble formation is an essential mechanism for the regulation and control of gene transcription
\cite{Polozov1988, Yakushevich2004}. Structure heterogeneities in the DNA chain and the presence of boundary interphase walls may affect the
propagation regime of these regulatory signals during replication processes.

In the scattering of waves by localised defects, a crucial point to consider is that nonlinear waves exhibit a highly complex behaviour. Transmission,
reflection, and annihilation are far from being the only possible phenomena that may occur. Indeed, solitons with an internal structure, such
as sine-Gordon (sG) kinks and breathers, may produce many remarkable phenomena when colliding with localised inhomogeneities. The main phenomenon is
the internal mode instability \cite{Gonzalez2002, Gonzalez2003}, which may cause the breaking of solitons, the formation of multikinks, the creation of
kink-antikink pairs and the formation of localised structures \cite{GarciaNustes2012, GarciaNustes2017}. The formation of such structures in sG
systems may have important implications not only in biological systems \cite{Ivancevic2013}, but also in physical systems such as
Josephson Junctions (JJs) \cite{Barone1982} and ferromagnetic materials \cite{Mikeska1978}.

In this article, we investigate the complex phenomena exhibited by a two-dimensional sG system with localised inhomogeneities. Formation of localised structures with bubble-like and drop-like shape are produced in two situations: (a) for line solitons trapped by a single localised
inhomogeneity, and (b) for travelling line solitons colliding with an array of localised inhomogeneities. The structure formation is understood using
the one-dimensional theory of activation of internal modes in sG solitons. The outline of the article is as follows: Section \ref{Sec:Theory} gives an overview of the one-dimensional theory of activation of internal modes of sG kinks under the action of inhomogeneous external forces. In Sec.
\ref{Sec:StationarySolitons} the two-dimensional model is introduced, and the internal mode instabilities of line solitons are studied. To avoid any scattering effect, we
consider steady line solitons. In Sec. \ref{Sec:Arrays}, the scattering of line solitons with a spatial
array of localised inhomogeneities is investigated. Finally, Sec. \ref{Sec:Conclusions} gives some concluding remarks.

\section{Preliminaries}
\label{Sec:Theory}

In this section, an overview of well-known preliminary results of sG kinks under the action of external forces is given. These results provide
the theoretical framework for the understanding of the phenomena reported in the subsequent sections.

\subsection{Kinks driven by homogeneous external forces}

The driven and damped two-dimensional sG system is a particular case of the more general nonlinear Klein-Gordon (KG) system
\begin{equation}
 \label{Eq01}
  \partial_{tt}\phi(\mathbf{r},t)-\nabla^2\phi(\mathbf{r},t)+\gamma\partial_t\phi(\mathbf{r},t)-G(\phi)=f,
\end{equation}
where $\mathbf{r}=(x,y)$, $\gamma$ is a linear-damping coefficient, $f$ is a constant external force and $G(\phi)=-\partial U(\phi)/\partial\phi$,
with the nonlinear potential $U(\phi)$ possessing at least two minima separated by a barrier.
It is well known that these two minima are fixed points of the
associated dynamical system, and that kink solutions are heteroclinic trajectories joining such fixed points \cite{vanSaarloos1990}. For sG systems,
the potential function is given by $U(\phi)=1-\cos\phi$ \cite{Peyrard2004}, and the equation (\ref{Eq01}) for $f=0$ has the well known stationary
kink-antikink solutions (line solitons), given by $ \phi_{\pm}(\mathbf{r})=4\arctan\exp\left(\pm x\right)$. Here, $\phi_+$ and $\phi_-$ reads
for the kink and antikink, respectively.

A space independent force with a constant value $f$ leads to an effective potential of the form $U_{eff}(\phi)=U(\phi)-f\phi$. In this case, kink
solutions exist for equation (\ref{Eq01}) only if this effective potential still have at least two minima separated by a barrier. That condition ceases to
be fulfilled for a certain critical value $f_c$ of the external force. When $f>|f_c|$ kink solutions do not exist, which means that an initially at
rest kink will be destroyed by the external force \cite{Gonzalez1992, Gonzalez2007}. The exact value of $f_c$ depends indeed on the particular
potential function of the system. For instance, using the sG potential, minima $\phi_j$ $(j=0,1\ldots)$ of the effective potential satisfy the
equation $\sin\phi_j=f$, which has no real solutions for $\phi_j$ if $|f|>1$. Thus, the critical value is $f_c=1$ for sG systems. 
 
If $f<|f_c|$, kink solutions in perturbed sG systems exists, and its dynamics in the neighbourhood of fixed points and separatrices can be
investigated using the so-called qualitative theory of dynamical systems \cite{Guckenheimer1986, McLaughlin1978, Kivshar1989, Sanchez1998,
Chacon2008, Gonzalez2007-II}. Based on this, it is possible to generalise the results to other equations that are topologically equivalent to those with the exact
solutions \cite{Gonzalez1996, Gonzalez2003}.

\subsection{Soliton breaking by inhomogeneous forces}

In the sense of the qualitative theory of dynamical systems, for a general $x$-dependent force $F(\mathbf{r})=f(x)$, it is known that a zero of
$f(x)$ at $x=x_*$ is an equilibrium point for the centre-of-mass position of the kink \cite{Gonzalez1992, Gonzalez1996, Gonzalez2003, Gonzalez2008}. The
equilibrium is stable if $\left.df(x)/dx\right|_{x=x_*}>0$, and unstable if $\left.df(x)/dx\right|_{x=x_*}<0$. If the kink is on a stable
equilibrium point, the shape and position of the kink will be recovered for any initial perturbation \cite{Gonzalez2003}. On the contrary, if the
kink is in an unstable equilibrium position, it is stretched by the inhomogeneous force that is acting on its body in opposite directions.
The value $f_c$ is the limit of $|f(x)|$ that the kink can resist the stretching without being destroyed \cite{Gonzalez2007}. If $|f(x)|>f_c\forall x$, the force will destroy the line soliton.

Notwithstanding, there is a more subtle mechanism for the kink destruction when $|f(x)|>f_c$ only in a localised region in space. This mechanism
is the internal mode instability \cite{Gonzalez2007-II}. In one-dimensional systems,
the stability analysis of internal modes of sG kinks has been solved exactly \cite{Gonzalez2002, Gonzalez2003}. The procedure starts by introducing
the ansatz
\begin{equation}
 \label{Eq02}
\phi(x)=4\arctan\exp\left(\pm Bx\right), 
\end{equation}
where $B$ is a parameter that controls the width of the kink. Solving an inverse problem, one obtains that equation (\ref{Eq02}) is a solution of the 
one-dimensional sG equation if the external force is given by
\begin{equation}
 \label{Eq02Force}
F(x) = 2(B^2 -1)\mbox{sinh}(Bx)\mbox{sech}^2(Bx),
\end{equation}
which is an antisymmetric spatial function that vanishes exponentially for $x\to\pm\infty$. For increasing (decreasing) values
of $B$, the force is less (more) localised in space and has decreasing (increasing) extreme values. Thus, the parameter $B$ also controls the extreme
values and the extension of $F(x)$ in space.
Interactions of DNA special sites with ligands or perturbation of fluxons by dipole currents in JJs are typical examples of strong and local
external influences with the same topological form as equation (\ref{Eq02Force}) \cite{Polozov1988, Malomed2004}.
From the stability analysis follows that the first stable internal shape mode arises for $1/6<B^2<1/3$, while for $B^2<1/6$ many other internal modes can
appear. For $B^2<2/[\Lambda_*(\Lambda_*+1)]$, where $\Lambda_*=(5+\sqrt{17})/2$, the first internal mode becomes unstable, leading to soliton
breakups and shape instabilities \cite{Gonzalez1992,Gonzalez2002, Gonzalez2003}. These instabilities play a central role in the formation of
structures reported in the following sections.

\section{Stationary line solitons in the presence of a single localised inhomogeneity}
\label{Sec:StationarySolitons}

\begin{figure}[t]
 \begin{center}
  \scalebox{0.27}{\includegraphics{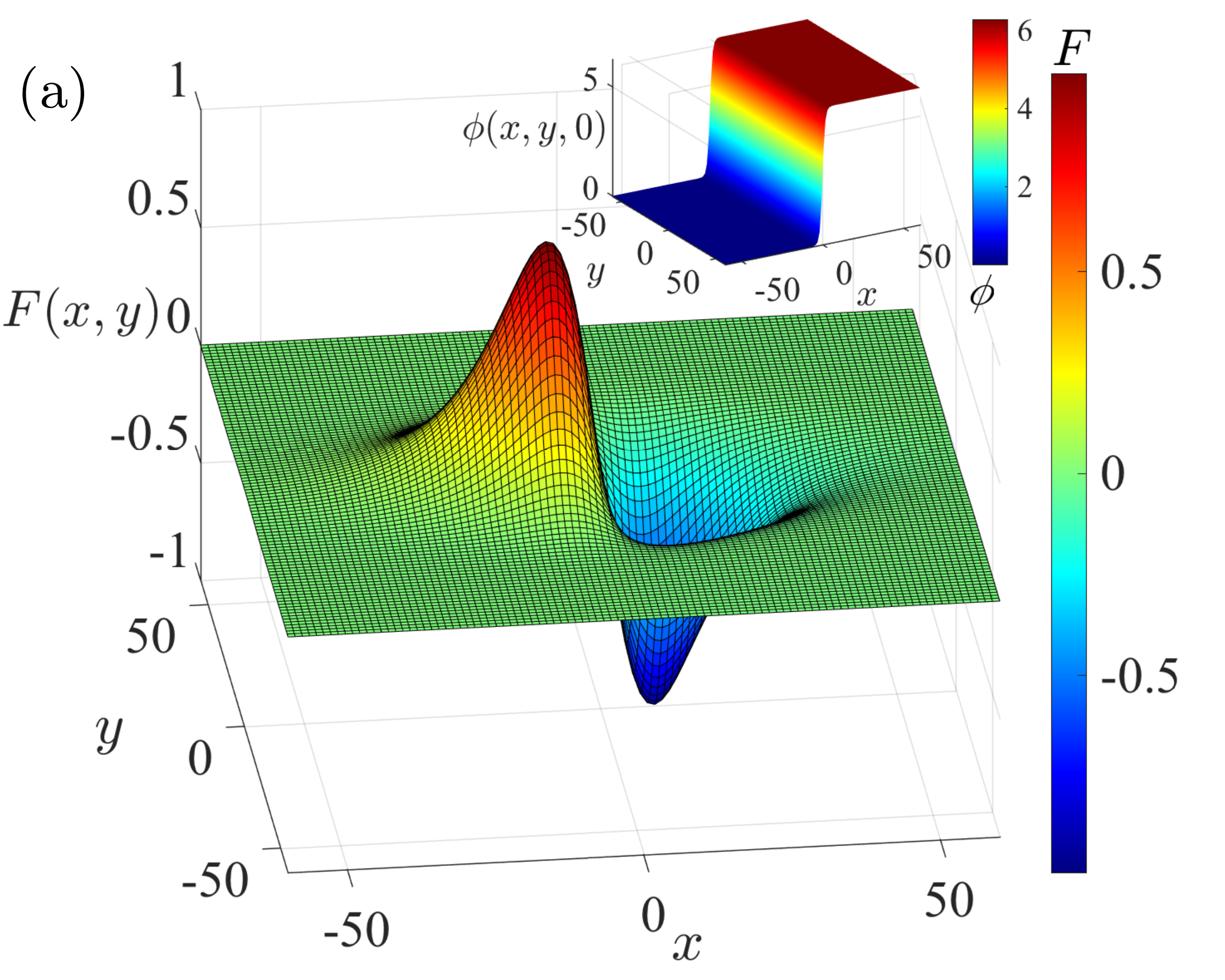}}\scalebox{0.27}{\includegraphics{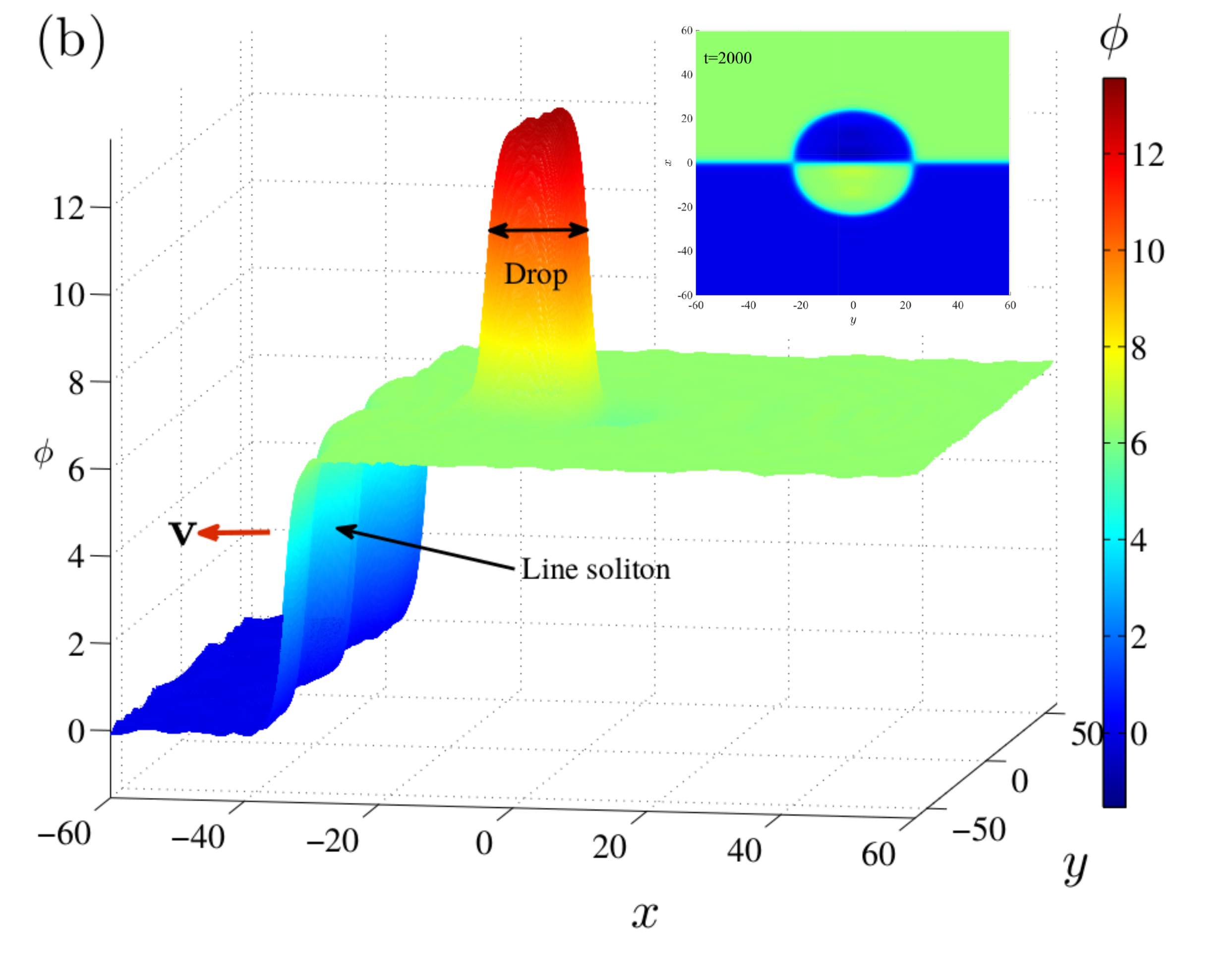}}\\
  \scalebox{0.27}{\includegraphics{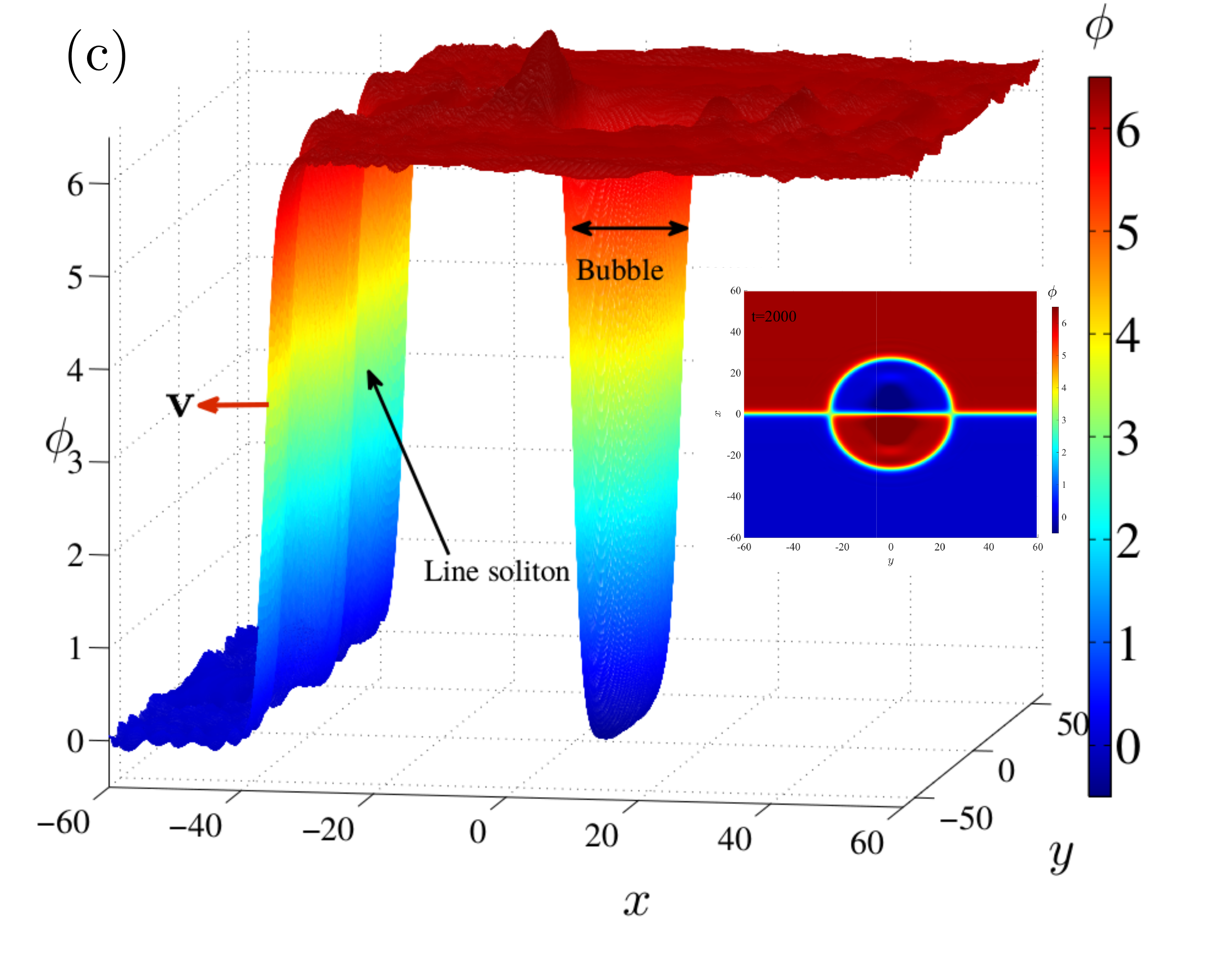}}\scalebox{0.27}{\includegraphics{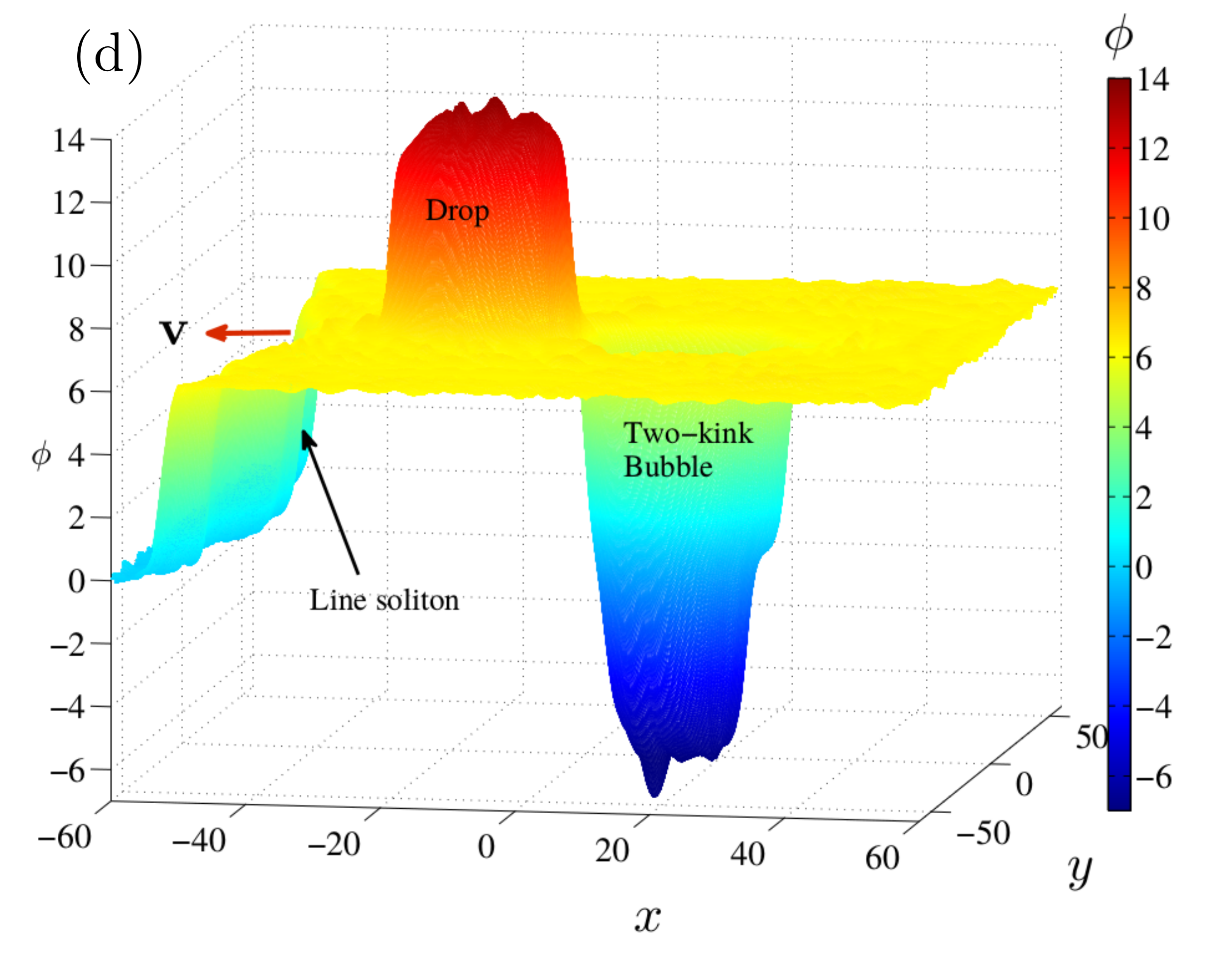}}
  \caption{(Colour online) (a) Inhomogeneous force of equation (\ref{Eq04}) for $B=0.1$ and $\sigma=15$. The inset shows the static line soliton
  used as initial condition in the numerical simulations of section \ref{Sec:StationarySolitons}. (b) A drop-like structure after $10\,000$ time
  iterations for $B=0.28$ and $\sigma=8.9$. The inset shows the elliptic-like structure that precedes the drop formation. The value of the field
  $\phi$ is indicated with the same colour scale. (c) A bubble-like structure after $10\,000$ time iterations for $B=0.27$ and $\sigma=10$.
  The inset shows the elliptic-like structure that precedes the bubble formation. (d) A multi-structure formed after $8\,300$ time iterations for
  $B=0.1$ and $\sigma=6$.}
  \label{fig01}
 \end{center}
\end{figure}

Following the one-dimensional system, consider the two-dimensional perturbed sG equation
\begin{equation}
 \label{Eq03}
 \partial_{tt}\phi(\mathbf{r},t)=\nabla^2\phi(\mathbf{r},t)-\sin\phi(\mathbf{r},t)-\gamma\partial_t\phi(\mathbf{r}, t)+F(\mathbf{r}),
\end{equation}
where the inhomogeneous force $F(x,y)$ is given by \cite{GarciaNustes2017}
\begin{equation}
\label{Eq04}
F(x, y)=2(B^2 -1)\mbox{sech}(Bx)\mbox{tanh}(Bx)e^{-y^2/\sigma^2},
\end{equation}
and a real parameter $\sigma$ has been introduced to control the spread of the force in the $y$-direction.
Figure \ref{fig01}(a) shows the force of equation (\ref{Eq04}), which represents a single localised inhomogeneity centred at the origin. The point
$x_*=0$ will be an equilibrium point for two-dimensional kinks. The external force of equation (\ref{Eq04}) is physically realisable in real
experiments. Indeed, the sG equation (\ref{Eq03}) with this kind of external perturbation models the propagation of fluxons in two-dimensional JJs
with a current dipole device inserted into one of the superconducting electrodes \cite{Ustinov2002, Malomed2004, GarciaNustes2017}.

This section shows how the inhomogeneity of equation (\ref{Eq04}) affects the dynamics of a line soliton, having a special focus on the dynamics of
the soliton internal structure. To avoid any scattering effect, consider an initially at rest kink with its centre-of-mass located at the equilibrium point. This initial condition is depicted in the inset of figure \ref{fig01}(a) and is given by
$\phi(x,y,0)=4\arctan\,\exp(x/L)$ with $\partial_t{\phi}(x,y,0)=0$, where $L$ is an arbitrary parameter.
We have used an explicit finite difference scheme of the second order of
accuracy for the numerical solution of equation (\ref{Eq03}), with $\gamma=0.01$ and homogeneous Neumann boundary conditions. The used regular mesh size is $1\,200\times 1\,200$. The space intervals are $\Delta x=\Delta y=0.1$, and the time increment is \hbox{$\Delta t=0.02$}.

Figure \ref{fig01} shows the main results obtained from numerical simulations.
Figure \ref{fig01}(b) shows a drop-like structure created for $B=0.28$ and $\sigma=8.9$. The local instability of the shape mode leads to a
local break-up of the initially at rest line soliton. This initial breaking consists in a local creation of a pair kink-antikink, giving an
elliptic-like structure that grows in time [see the inset of figure \ref{fig01}(b)]. This kink zone propagates until the localised force is
weak enough, and the contribution of curvature effects and kink interactions are important. The elliptic form eventually collapses due to the attractive interaction forces between the kink and
antikink, developing a final stable drop-like structure in the negative region of $x$. Notice that the drop is sustained by the positive part of the external force. A kink is still present in the system due to the conservation of the topological charge and propagates
towards the negative $x$-direction with a velocity $\mathbf{v}$, as showed in figure \ref{fig01}(b).

If we proceed to diminish the value of $B$, the preferred structure is one with the reverse topological charge: a bubble-like
structure. Figure \ref{fig01}(c) shows a bubble-like structure created for $B=0.27$ and $\sigma=10$. In this case, the force is extended enough in
the $x$-direction to prevent the collapse and sustain the walls of the elliptic form, forming a bubble sustained by the negative part of the force. The mechanism of bubble formation
is similar to the drop formation, giving that both are preceded by an elliptic-like form that grows in time [see inset of figure \ref{fig01}(c)].
Over a critical value of the elliptical area, the kink-antikink interaction is too weak to produce the collapse of the bubble. In this case, the external force supports the structure. Otherwise, the external force cannot sustain the walls of the bubble and the structure collapses by kink-antikink interaction, triggering the formation of a drop-like structure. Therefore, the interplay of both parameters $B$ and $\sigma$ will determine which kind of structure will be formed.
Indeed, further numerical simulations have show that these instabilities in sG kinks can be activated varying any of both parameters $B$ or $\sigma$
\cite{GarciaNustes2017}.

If we continue further diminishing the values of parameters $B$ and $\sigma$, it is expected that the line soliton will exhibit more complex
phenomena due to the local destabilisation of more internal modes. Indeed, there is a coexistence region in the parameter space where a bubble
and a drop merge in a stable structure, like the bubble-drop bound state depicted in figure \ref{fig01}(d) for $B=0.1$ and $\sigma=6$. From the
one-dimensional theory, it is known that for $B\lesssim0.2564$ the formation of two-kink solitons occurs \cite{GarciaNustes2012}. The later explains the
two-kink profile of the bubble showed in figure \ref{fig01}(d). More complex multi-structures can be observed for smaller values of $B$ and
$\sigma$, such as multi-kink bubbles and drops.

Notice that the localised structures showed in Fig.~\ref{fig01} are called \emph{bubbles} and \emph{drops} giving that the equation (\ref{Eq03}) also appears in the description of structural phase transitions \cite{GarciaNustes2017, Gonzalez1999, Gonzalez2006}. The results of this article can
be interpreted as the driving of a system to a new phase by the activation of a local instability.

\section{Scattering of solitons in an array of localised inhomogeneities}
\label{Sec:Arrays}

In order to simulate some real situations in physical and biological systems, this section deals with travelling line solitons that collide with a
distribution of localised inhomogeneities. Consider in equation (\ref{Eq03}) the following spatial distribution of inhomogeneous forces
\begin{equation}
 \label{Eq05}
F(x,y)= F_o+2\sum_{i=1}^4(B_i^2-1)\mbox{sech}\left[B_i(x-x_i)\right]\mbox{tanh}\left[B_i(x-x_i)\right]e^{-(y-y_i)^2/\sigma^2},
\end{equation}
where $F_o=0.1$ is a constant homogeneous contribution to the total external force, $(x_i,y_i)$ is the coordinate-position of the core of the $i$-th inhomogeneity, $B_i$ is its
respective control parameter, $\sigma=10$, and $i\in\left\{1,\,2,\,3,\,4\right\}$. Notice that the inhomogeneous contribution of force (\ref{Eq05})
is given by the superposition of forces of the same form as equation (\ref{Eq04}).

The position of the inhomogeneities are fixed for all simulations in this section, and are given by $\mathbf{r}_1=(-15, 25)$, $\mathbf{r}_2=(15, 25)$, $\mathbf{r}_3=(-15, -25)$ and
$\mathbf{r}_4=(15, -25)$. With these coordinates, the inhomogeneities are placed at the vertices of a rectangle centred at the origin.
The initial condition is given by $\phi(x,y,0)=4\arctan\,\exp[(x-xo)/L]$ with $\partial_t\phi(x,y,0)=0$, which is an initially at rest line soliton
placed some distance away from the array of inhomogeneities. The initial $x$-coordinate of the centre-of-mass of the kink is given by $x_o=50$. Due to the homogeneous part of the force (\ref{Eq05}) and the damping term of equation (\ref{Eq03}), the line soliton accelerates towards the negative $x$-direction before reaching an asymptotically constant velocity. When the soliton collides with the inhomogeneities, many remarkable phenomena may occur due to possible internal mode
instabilities.

\begin{figure}[t]
 \begin{center}
  \scalebox{0.32}{\includegraphics{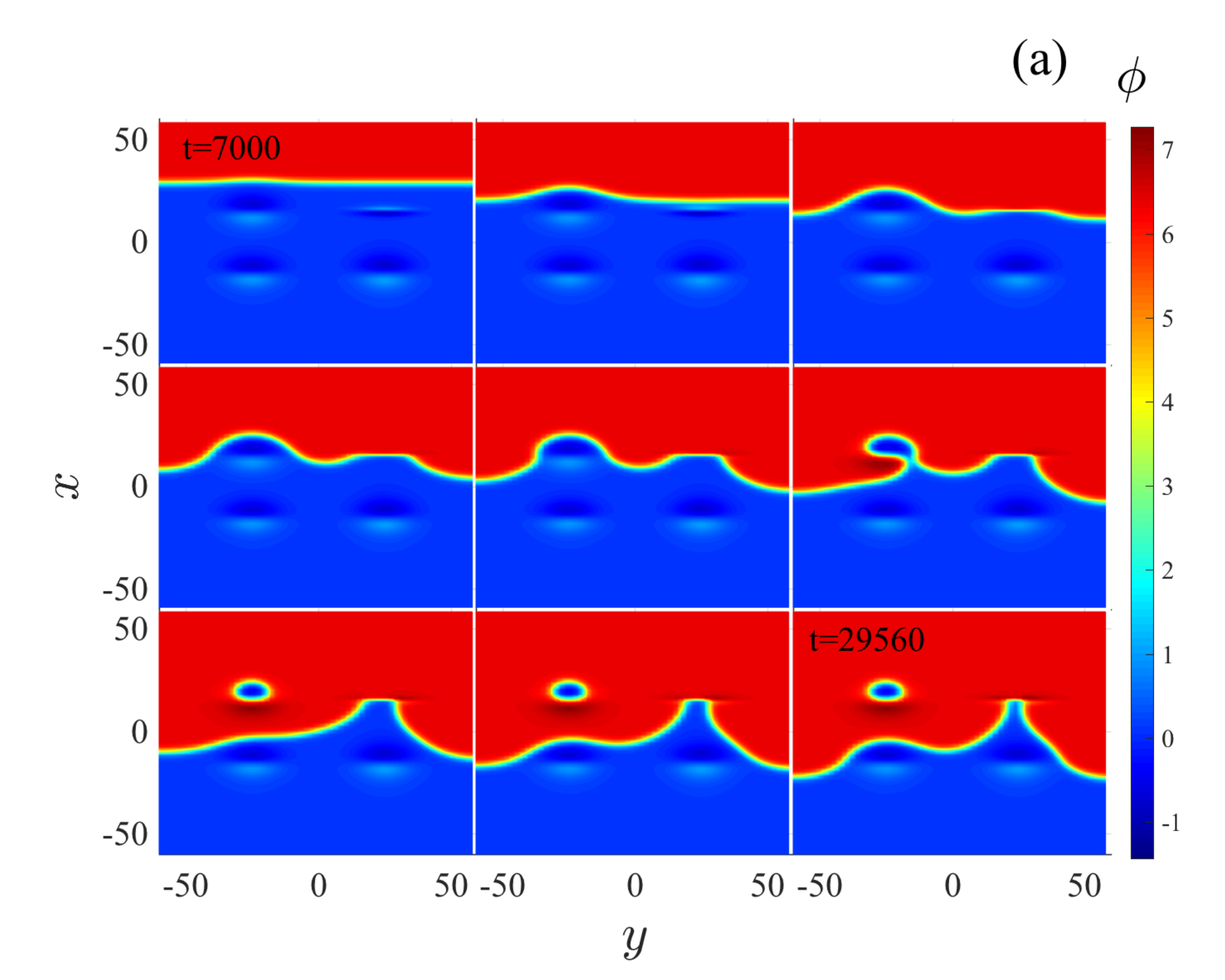}}\scalebox{0.32}{\includegraphics{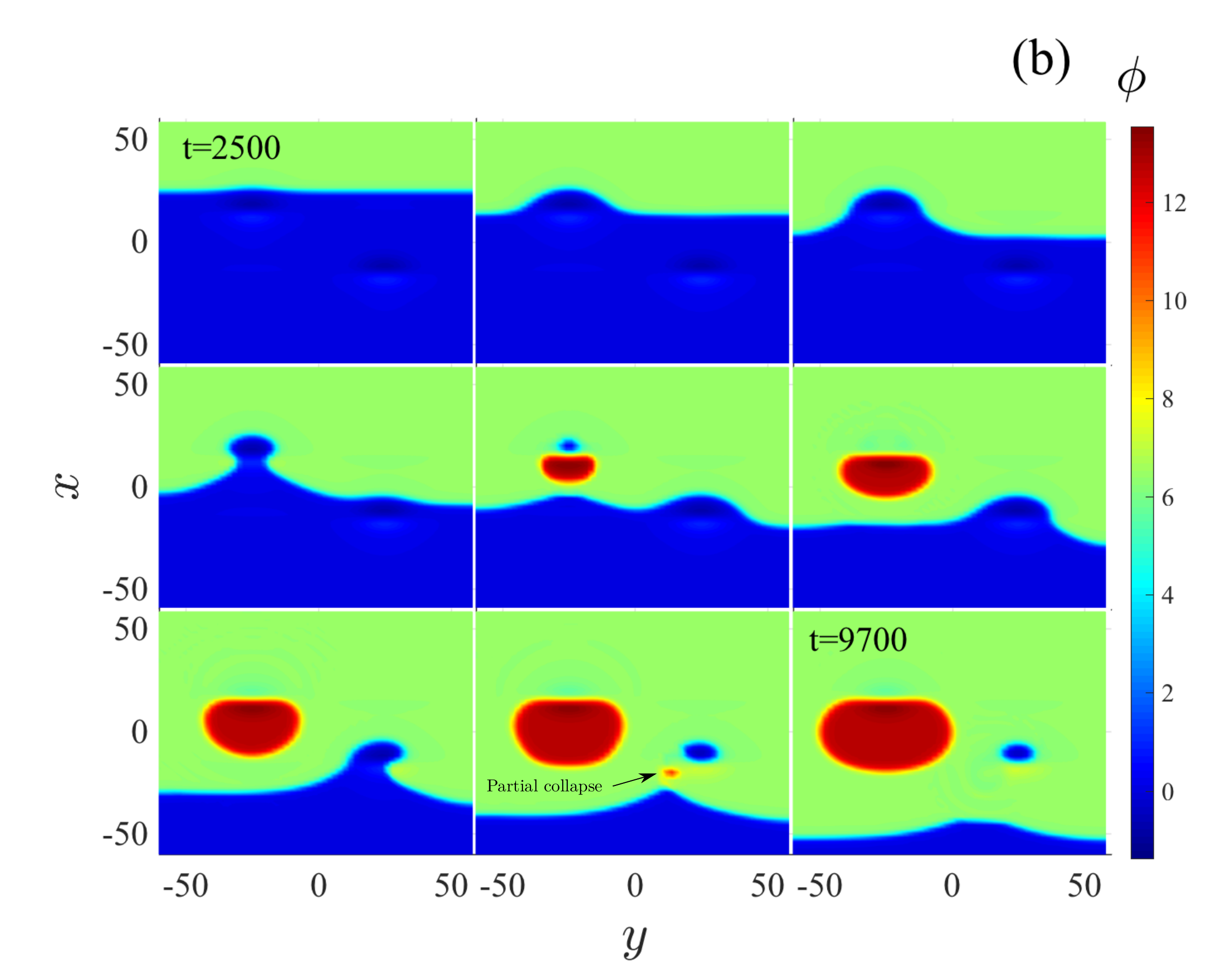}}\\
  \scalebox{0.32}{\includegraphics{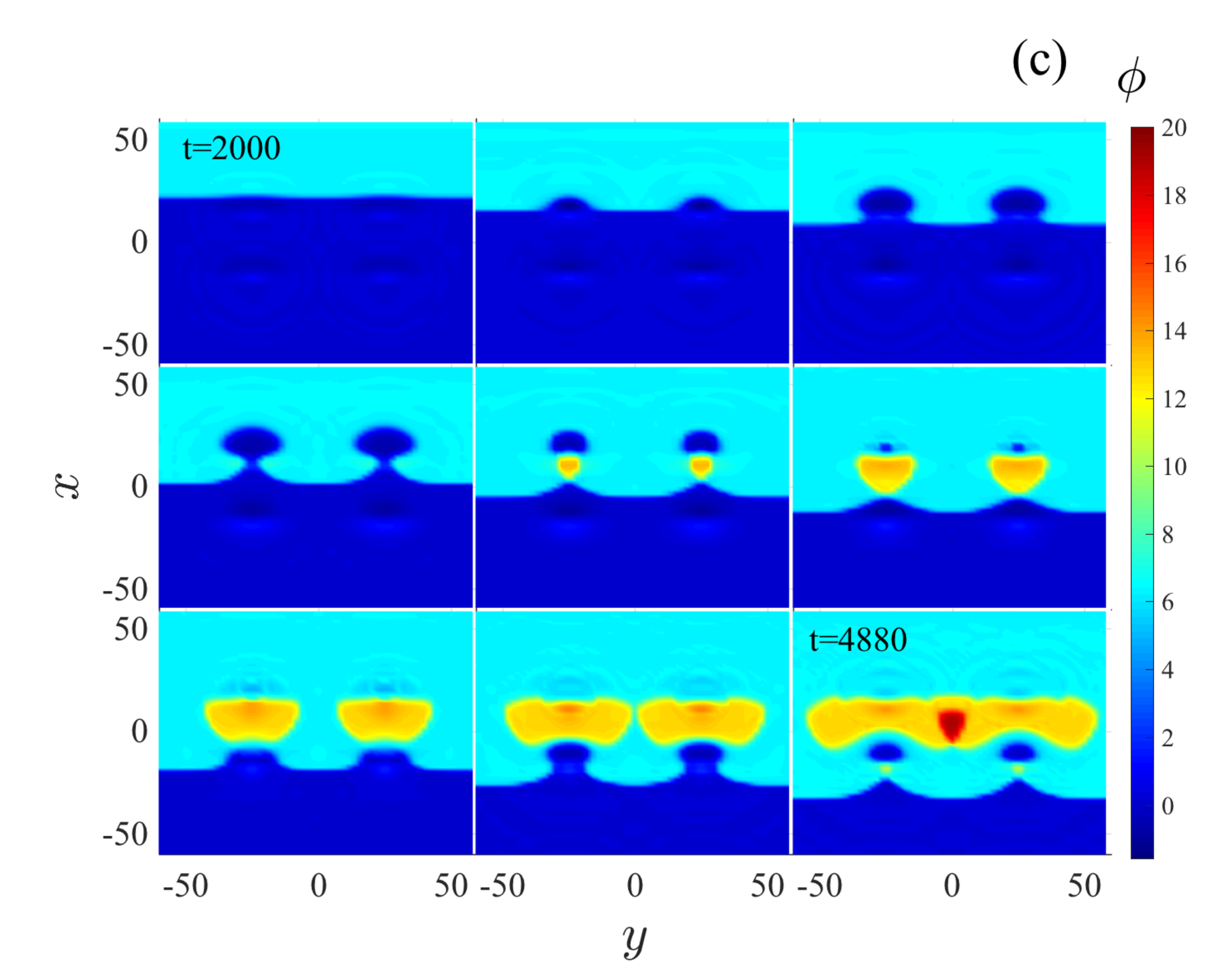}}
  \caption{(Colour online) Scattering of travelling line solitons after collisions with a distribution of localised inhomogeneities. (a) Generation
  of a bubble induced by the collapse of an echo-wave, for $B_1=0.27$, $B_2=3.0$, $B_3=B_4=0.3$, and $\gamma=0.5$. (b) Kink transmission for
  $\gamma=0.1$, $B_1=B_4=0.27$ and $B_2=B_3=1.1$. A bubble and a drop remains trapped at the inhomogeneities after the transmission of the soliton.
  (c) Kink transmission and formation of complex interacting structures (bubbles and drops) for $\sigma=10$, $\gamma=0.01$, and
  $B_i=0.27$ $(i=1,2,3,4)$.}
  \label{fig02}
 \end{center}
\end{figure}

Figure \ref{fig02}(a) shows the results from numerical simulations for $B_1=0.27$, $B_2=3.0$, $B_3=B_4=0.3$, and $\gamma=0.5$. For this combination of
parameters, the $1$st inhomogeneity activates the internal modes of kinks, while the $2$nd inhomogeneity tends to trap them at position
$\mathbf{r}_2$. When the line soliton reaches the first line of inhomogeneities, part of the kink is deformed due to the collision with the $1$st
inhomogeneity, while another part is trapped by the $2$nd inhomogeneity. Notice that this initial bent of the soliton is the so-called \emph{echo effect} reported in
biophysical systems and is considered as responsible for certain types of disorders and cardiac arrhythmias \cite{Mornev2000}. This echo-wave is eventually stabilised in size, and forms a stable bubble-like structure sustained by the $1$st inhomogeneity. Notice that the echo-waves observed
in this section have some resemblance with the elliptic-like structures observed in section \ref{Sec:StationarySolitons} [see insets of figure
\ref{fig01}(b) and (c)].

For smaller values of $\gamma$ the soliton reaches the inhomogeneities with higher terminal speeds, and we would expect the
enhancing of the transmission of the soliton across the inhomogeneities. Figure \ref{fig02}(b) shows the results from numerical simulations for $\gamma=0.1$, $B_1=B_4=0.27$ and
$B_2=B_3=1.1$. In this case, the $1$st and $4$th inhomogeneities activate the internal modes of the soliton. The $2$nd and $3$rd
inhomogeneities tend weakly to trap the soliton. Indeed, when the kink collides with the first pair of inhomogeneities, the $2$nd cannot capture the
kink, while the $1$st inhomogeneity produces an echo-effect. This echo-wave eventually collapse and forms a stable drop, as showed in figure
\ref{fig02}(b). A similar mechanism occurs when the line soliton reaches the second pair of inhomogeneities. However, the speed of the line
soliton has diminished due to previous collisions. The echo-wave generated at the $4$th inhomogeneity partially collapse, as indicated in figure
\ref{fig02}(b), but is eventually stabilised and forms a bubble. Thus, the relative speed of collision also plays an important role in determining
which kind of structure will be formed. 

A highly complex dynamics is observed diminishing the parameters $B_i$ in the array of inhomogeneities. Figure \ref{fig02}(c) shows the
results from numerical simulations when $B_i=0.27$ for all $i$. After the collision, there is an echo effect followed by the formation of bubbles.
The inhomogeneities cannot sustain such bubbles, and they collapse to form drops eventually. However, during this drop formation, another pair of
bubbles are created in the neighbour inhomogeneities. In figure \ref{fig02}(c) is clearly seen repulsive interactions between these new bubbles and
the previously formed drops. Notice that these drops also collide, enhancing the formation of new structures with the topological charge of three around
the origin. More complex dynamics can be observed diminishing the values of $B_i$ or $\sigma$.

\section{Conclusion}
\label{Sec:Conclusions}

In this article, the formation of localised structures in a two-dimensional sG system has been reported considering two different scenarios: (a) the
presence of a single localised inhomogeneity, and (b) the presence of an array of localised inhomogeneities. These inhomogeneities are modelled by a family of topologically equivalent external forces. The localised structures are formed with a bubble-like and drop-like shape and are
sustained by the same inhomogeneities that create them. The structure creation is entirely due to the dynamics of the internal modes of the soliton,
and not to any scattering effects. However, the internal mode instabilities can also occur during scattering processes, and exhibit a highly complex
dynamics.

As a final remark, notice that the reported phenomena may have important implications in biological systems. The formation of bubbles, and the 
trapping and breaking of such structures by localised inhomogeneities may be related to different functional disorders in similar systems.
Moreover, the study of the stability and transport of such bubbles may give some light on how to introduce shock perturbations, and produce the
desired effects in a controlled way. Gene transcription in DNA chains can also be regulated by the insertion of this kind of external perturbations.

\ack

This article was written after a conference at the XX Symposium of the Physical Society of Chile (SOCHIFI), held in Santiago de Chile on November 2016.
The author thanks the organising committee for this occasion. The author also thanks Prof. M\'onica A. Garc\'ia-\~Nustes and Prof. Jorge A.
Gonz\'alez for many fruitful discussions and the active collaboration on the subject of this article. This work was supported by CONICYT doctorado nacional No.
21150292.

\section*{References}

\begin{thebibliography}{10}
\expandafter\ifx\csname url\endcsname\relax
  \def\url#1{{\tt #1}}\fi
\expandafter\ifx\csname urlprefix\endcsname\relax\def\urlprefix{URL }\fi
\providecommand{\eprint}[2][]{\url{#2}}

\bibitem{Ivancevic2013}
Ivancevic V and Ivancevic T 2013 {\em J. Geom. Symmetry Phys.\/} {\bf 31} 1--56

\bibitem{Aslanidi1999}
Aslanidi O~V and Mornev O~A 1999 {\em J. Biol. Phys.\/} {\bf 25} 149--164 ISSN
  0092-0606

\bibitem{Mornev2000}
Mornev O~A, Tsyganov L~M, Aslanidi O, Ordanovich A~E and Chailakhyan L~M 2000
  {\em Dokl. Biophys.\/} {\bf 373} 32--36 ISSN 0012-4974

\bibitem{Scott1975}
Scott A~C 1975 {\em Rev. Mod. Phys.\/} {\bf 47}(2) 487--533

\bibitem{Polozov1988}
Polozov R~V and Yakushevich L~V 1988 {\em J. Theor. Biol.\/} {\bf 130} 423--430
  ISSN 0022-5193

\bibitem{Yakushevich2004}
Yakushevich L~V 2004 {\em Nonlinear Physics of DNA\/} (Weinheim: Wiley--VCH)

\bibitem{Gojkovic2016}
Gojkovic Z and Ivancevic T 2016 {\em Nonlinear Dynamics\/} {\bf 86} 2071--2080
  ISSN 1573-269X

\bibitem{Grinevich2015}
Grinevich A~A, Ryasik A~A and Yakushevich L~V 2015 {\em Chaos Soliton.
  Fract.\/} {\bf 75} 62--75 ISSN 0960-0779

\bibitem{Grinevich2016}
Grinevich A~A and Yakushevich L~V 2016 {\em Biophysics\/} {\bf 61} 539--546
  ISSN 1555-6654

\bibitem{Gonzalez2002}
Gonz\'alez J~A, Bellor\'in A and Guerrero L~E 2002 {\em Phys. Rev. E\/} {\bf
  65}(6) 065601

\bibitem{Gonzalez2003}
Gonz\'alez J~A, Bellor\'in A and Guerrero L~E 2003 {\em Chaos Soliton.
  Fract.\/}  907--919 ISSN 0960-0779

\bibitem{GarciaNustes2012}
Garc\'{\i}a-\~Nustes M~A and Gonz\'alez J~A 2012 {\em Phys. Rev. E\/} {\bf
  86}(6) 066602

\bibitem{GarciaNustes2017}
Garc\'{\i}a-\~Nustes M~A, Mar\'{\i}n J~F and Gonz\'alez J~A 2017 {\em Phys.
  Rev. E\/} {\bf 95}(3) 032222

\bibitem{Barone1982}
Barone A and Patern{\`o} G 1982 {\em Physics and Applications of the Josephson
  Effect\/} (New York: Wiley)

\bibitem{Mikeska1978}
Mikeska H~J 1978 {\em J. Phys. C: Solid State Phys.\/} {\bf 11} L29--32

\bibitem{vanSaarloos1990}
van Saarloos W and Hohenberg P~C 1990 {\em Phys. Rev. Lett.\/} {\bf 64}(7)
  749--752

\bibitem{Peyrard2004}
Peyrard M and Dauxois T 2004 {\em Physique des Solitons\/} (Paris: Savoirs
  Actuels)

\bibitem{Gonzalez1992}
Gonz\'alez J~A and Holyst J~A 1992 {\em Phys. Rev. B\/} {\bf 45}(18)
  10338--10343

\bibitem{Gonzalez2007}
Gonz\'alez J, Bellor\'in A and Guerrero L 2007 {\em Chaos Soliton. Fract.\/}
  {\bf 33} 143--155 ISSN 0960-0779

\bibitem{Guckenheimer1986}
Guckenheimer J and Holmes P 1986 {\em Nonlinear Oscillations, Dynamical
  Systems, and Bifurcations of Vector Fields\/} (New York: Springer-Verlag)

\bibitem{McLaughlin1978}
McLaughlin D~W and Scott A~C 1978 {\em Phys. Rev. A\/} {\bf 18}(4) 1652--1680

\bibitem{Kivshar1989}
Kivshar Y~S and Malomed B~A 1989 {\em Rev. Mod. Phys.\/} {\bf 61}(4) 763--915

\bibitem{Sanchez1998}
S\'anchez A and Bishop A~R 1998 {\em SIAM Rev.\/} {\bf 40} 579--615

\bibitem{Chacon2008}
Chac\'on R, Bellor\'{\i}n A, Guerrero L~E and Gonz\'alez J~A 2008 {\em Phys.
  Rev. E\/} {\bf 77}(4) 046212

\bibitem{Gonzalez2007-II}
Gonz\'alez J~A, Cuenda S and S\'anchez A 2007 {\em Phys. Rev. E\/} {\bf 75}(3)
  036611

\bibitem{Gonzalez1996}
Gonz\'alez J~A and Mello B~A 1996 {\em Phys. Scr.\/} {\bf 54} 14

\bibitem{Gonzalez2008}
Gonz\'alez J~A, Garc\'ia-\~Nustes M~A, S\'anchez A and McClintock P~V~E 2008
  {\em New J. Phys.\/} {\bf 10} 113015

\bibitem{Malomed2004}
Malomed B~A and Ustinov A~V 2004 {\em Phys. Rev. B\/} {\bf 69}(6) 064502

\bibitem{Ustinov2002}
Ustinov A~V 2002 {\em Appl. Phys. Lett.\/} {\bf 80} 3153--3155

\bibitem{Gonzalez1999}
Gonz\'alez J~A and Oliveira F~A 1999 {\em Phys. Rev. B\/} {\bf 59}(9)
  6100--6105

\bibitem{Gonzalez2006}
Gonz\'alez J, Marcano A, Mello B and Trujillo L 2006 {\em Chaos Soliton.
  Fract.\/} {\bf 28} 804--821 ISSN 0960-0779

\end{thebibliography}

\providecommand{\noopsort}[1]{}\providecommand{\singleletter}[1]{#1}%
\providecommand{\newblock}{}


\end{document}